\documentclass[twocolumn]{aastex62} 
\usepackage{subfigure}
\usepackage{url}
\usepackage{graphicx,color}
\usepackage{gensymb}     

\usepackage{hyperref}
\newcommand{\CII}{\ion{C}{2}}

\newcommand{\CaII}{\ion{Ca}{2}}
\newcommand{\MgII}{\ion{Mg}{2}}

\newcommand{\NaI}{\ion{Na}{1}}

\DeclareMathAlphabet{\mathitbf}{OML}{cmm}{b}{it}
\DeclareMathAlphabet{\mathf}{OML}{cmm}{c}{sl}

\newcommand{\eg}{e.g.,}

\newcommand{\kms}{km~s$^{-1}$}

\shorttitle{Twisting and Photospheric Mixed-polarity Field in Large Penumbral Jets}
\shortauthors{Tiwari et al.}

\begin{document}

\title{Evidence of Twisting and Mixed-polarity Solar Photospheric Magnetic Field in Large Penumbral Jets: IRIS and Hinode Observations}


\correspondingauthor{Sanjiv K. Tiwari}
\email{tiwari@lmsal.com}

\author[0000-0001-7817-2978]{Sanjiv K. Tiwari}
\affiliation{Lockheed Martin Solar and Astrophysics Laboratory, 3251 Hanover Street, Bldg. 252, Palo Alto, CA 94304, USA}
\affiliation{Bay Area Environmental Research Institute, NASA Research Park, Moffett Field, CA 94035, USA}

\author{Ronald L. Moore}
\affiliation{NASA Marshall Space Flight Center, Mail Code ST 13, Huntsville, AL 35812, USA}
\affil{Center for Space and Aeronomic Research, The University of Alabama in Huntsville, Huntsville, AL 35805, USA}

\author{Bart De Pontieu}
\affiliation{Lockheed Martin Solar and Astrophysics Laboratory, 3251 Hanover Street, Bldg. 252, Palo Alto, CA 94304, USA}
\affil{Rosseland Centre for Solar Physics, University of Oslo, P.O. Box 1029 Blindern, NO–0315 Oslo, Norway}
\affil{Institute of Theoretical Astrophysics, University of Oslo, P.O. Box 1029 Blindern, NO–0315 Oslo, Norway}

\author{Theodore D. Tarbell}
\affiliation{Lockheed Martin Solar and Astrophysics Laboratory, 3251 Hanover Street, Bldg. 252, Palo Alto, CA 94304, USA}

\author{Navdeep K. Panesar}
\affiliation{NASA Marshall Space Flight Center, Mail Code ST 13, Huntsville, AL 35812, USA}

\author{Amy R. Winebarger}
\affiliation{NASA Marshall Space Flight Center, Mail Code ST 13, Huntsville, AL 35812, USA}

\author{Alphonse C. Sterling}
\affiliation{NASA Marshall Space Flight Center, Mail Code ST 13, Huntsville, AL 35812, USA}


\graphicspath{{Figures/}}


\begin{abstract}

A recent study using {\it Hinode} (SOT/FG) data of a sunspot revealed some unusually large penumbral jets that often repeatedly occurred at the same locations in the penumbra, namely at the tail of a penumbral filament or where the tails of multiple penumbral filaments converged. These locations had obvious photospheric mixed-polarity magnetic flux in \NaI\ 5896 Stokes-V images obtained with SOT/FG. Several other recent investigations have found that extreme ultraviolet (EUV)/X-ray coronal jets in quiet Sun regions (QRs), coronal holes (CHs) and near active regions (ARs) have obvious mixed-polarity fluxes at their base, and that magnetic flux cancellation prepares and triggers a minifilament flux-rope eruption that drives the jet. Typical QR, CH, and AR coronal jets are up to a hundred times bigger than large penumbral jets, and in EUV/X-ray images show clear twisting motion in their spires. Here, using IRIS \MgII\ k 2796 \AA\ SJ images and spectra in the penumbrae of two sunspots we characterize large penumbral jets. We find redshift and blueshift next to each other across several large penumbral jets, and interpret these as untwisting of the magnetic field in the jet spire. Using Hinode/SOT (FG and SP) data, we also find mixed-polarity magnetic flux at the base of these jets. Because large penumbral jets have mixed-polarity field at their base and have twisting motion in their spires, they might be driven the same way as QR, CH and AR coronal jets.

\end{abstract}


\section{Introduction} \label{sec:intr}

Penumbral jets are small-scale transient events continuously occurring in sunspot penumbra, first detected by \cite{kats07} in movies taken in the chromospheric line \CaII\ H by {\it Hinode} \citep[Solar Optical Telescope/Filtergraph: SOT/FG,][]{kosu07,tsun08,ichi08}. These penumbral jets, also referred as microjets, are narrow (have a width of less than 400 km), live less than a minute, and have speeds of more than 100 \kms\ \citep{kats07}.  Penumbral jets are more clearly visible in the limbward side of penumbra than the disk-center side of penumbra due to foreshortening effect. In a recent study, \cite{tiw16} found penumbral jets that were larger than microjets, and repeatedly occurred at the tail of a penumbral filament (the ``tail" is the part of a filament that is farthest from the sunspot umbra) or where the tails of several penumbral filaments evidently converged. These jets were called as ``large" or ``tail" penumbral jets. Large penumbral jets were found to be much wider than microjets (up to 600 km when measured using FWHM of a Gaussian fit, or up to 1500 km when measured directly from the intensity enhancements as compared to the background i.e., using the Gaussian width), can have speeds of more than 200 \kms, and live as long as or often longer than microjets.  

Penumbral microjets were originally proposed to form due to component magnetic reconnection between two inclined fields of the same magnetic polarity \citep{kats07}. \cite{tiw16} proposed a modified picture of formation of penumbral microjets based on the recent observation of the internal structure of penumbral filaments and spines. See \cite{tiw17sunspot} for a recent review on sunspot structure. The penumbra is formed entirely of ``spines" and ``filaments" \citep{tiw15aa}. Spines \citep{lites93,titl93} are the intrusions of umbral field into the penumbra (in-between penumbral filaments), and is thus nearly vertical field, the inclination of which with respect to vertical increases with radial distance from the center of sunspots \citep{tiw15aa}. Penumbral filaments are elongated magnetized convective cells \citep{remp12,tiw13}: these have strong upflows (of $\sim$5 \kms) near the bright head (the ``head" is the part of a filament that is nearest to the sunspot umbra) with more nearly vertical and strong field (of 1.5 - 2 kG), and the upflow continues along the central axis of the filament up to more than half of its length, along the field having strength $\sim$1 kG. The tails of penumbral filaments contain strong downflows (of $\sim$7 \kms) and strong field (of 2 - 3.5 kG); they are dark in white light images and have opposite polarity field to that of heads. 

In addition to the downflows in the filament tails, less-intense downflows also occur along
the edges of penumbral filaments, and the magnetic polarity along these edges is often opposite to that of the
head, the spine or the umbra \citep[see also,][]{josh11,scha13,ruiz13,esta15,josh17}. Observe that, in this geometry, the field in spines and the field in the sides of filaments are both in (i) close proximity to each other and (ii) magnetically oriented in opposite directions, and thus are prone to magnetic reconnection. 

In the new picture presented by \cite{tiw16}, penumbral microjets are produced by magnetic reconnection between the spine field and the opposite-polarity field accompanied by the downflows in the sides of penumbral filaments, and propagate along the spine field. This picture is in partial agreement with the numerical magnetohydrodynamic (MHD) modelling of \cite{saka08} and \cite{maga10}. 


To validate the new picture of the formation mechanism of penumbral microjets, proposed by \cite{tiw16}, we need to establish one-to-one correspondence of the locations of penumbral microjets in the chromosphere to the photosphere, which is not trivial \citep{jurc08,tiw16}. Advanced processing techniques were used on different ground and space based data to detect the downflows and accompanying opposite-polarity magnetic field in the sides of penumbral filaments \citep{tiw13,van13,ruiz13,scha13,josh17}. Nonetheless, in a recent detailed study, observational support for the new picture of the formation mechanism of penumbral microjets has been found by \cite{este18}. 


Large penumbral jets have obvious mixed-polarity magnetic field in the photosphere \citep{tiw16}, consistent with them being located at the tails of penumbral filaments, which have opposite polarity field to that of the spines, umbral or filament-head field \citep{tiw13}. It was suggested by \cite{tiw16} that there might not be any fundamental differences in the formation mechanisms of penumbral microjets and large penumbral jets -- both form due to magnetic reconnection in the (higher) photosphere leading to a flux cancellation at the base. The only differences might lie in their sizes and contents of magnetic flux at their bases -- large  penumbral jets have larger amount of opposite-polarity flux, thus, larger jets are produced repeatedly until the opposite-polarity flux is completely cancelled, whereas the opposite-polarity field at the base of microjets, in the sides of filaments, is quite narrow and small, and is rarely visible \citep{tiw13}. 

\cite{tiw16} found clear signatures of large penumbral jets in the transition region (using AIA 1600, 171, 193 \AA\ data), consistent with the results of \cite{viss15}, who found transition region signatures of microjets using IRIS data. Because \cite{viss15} did not differentiate between microjets and large penumbral jets, we suspect that they probably only studied the largest of microjets, or perhaps some large penumbral jets. But even the largest of the penumbral jets studied by \cite{tiw16} never showed up in the corona (i.e., in AIA 94 \AA\ images), thus whether they have any direct contribution to coronal heating above sunspots remains unknown. 

Solar coronal jets at much larger scales (a hundred times bigger) than large penumbral jets have been extensively studied in coronal holes (CHs), quiet regions (QRs), and near active regions (ARs), both observationally and by numerical modelling \citep[e.g.,][]{shib92,cirt07,moor10,schm13,pari15,ster15,cheu15}. These coronal jets have been found to have helical or twisting motions in their spires, and to occur at the locations of mixed-polarity field in the photosphere, often with flux cancellation \citep{pats08,liu09,schm13,moor15,cheu15,pane16homo,pane16qr,ster17,pane18}.


Because large penumbral jets have been found to have mixed-polarity field at their base in the photosphere, in the present work, we focus on searching for any evidence of twisting in them, which would suggest similarities in the origin of penumbral jets and coronal jets. Jets in CHs, QRs and ARs are much bigger, thus, twisting in them is often clearly discernible in EUV intensity images. However, large penumbral jets are much smaller and it is not possible to detect twisting in direct intensity images. We therefore use IRIS \MgII\ spectra to search for possible blueshifts and redshifts compatible with twisting in large penumbral jets. We also verify the presence of mixed-polarity magnetic field at their base (and calculate flux cancellation rates), whenever the magnetic field data permits. 


\section{Data and Methods} \label{data}
We have selected two sunspots, both near solar disk center, whose penumbrae were observed nearly simultaneously by the Interface Region Imaging Spectrograph \cite[IRIS:][]{depo14IRIS}, and the Solar Optical Telescope \citep[SOT:][]{tsun08,ichi08,suem08,shim08,lites13SP} onboard Hinode \citep{kosu07}. 
From IRIS, we mainly focus on the \MgII\ line because its formation can be considered nearest to that of the \CaII\ line in which penumbral jets were first detected by \cite{kats07}.

\begin{figure*}
	\centering
	\includegraphics[trim=2.8cm 4.11cm 3.1cm 3.8cm,clip,width=\textwidth]{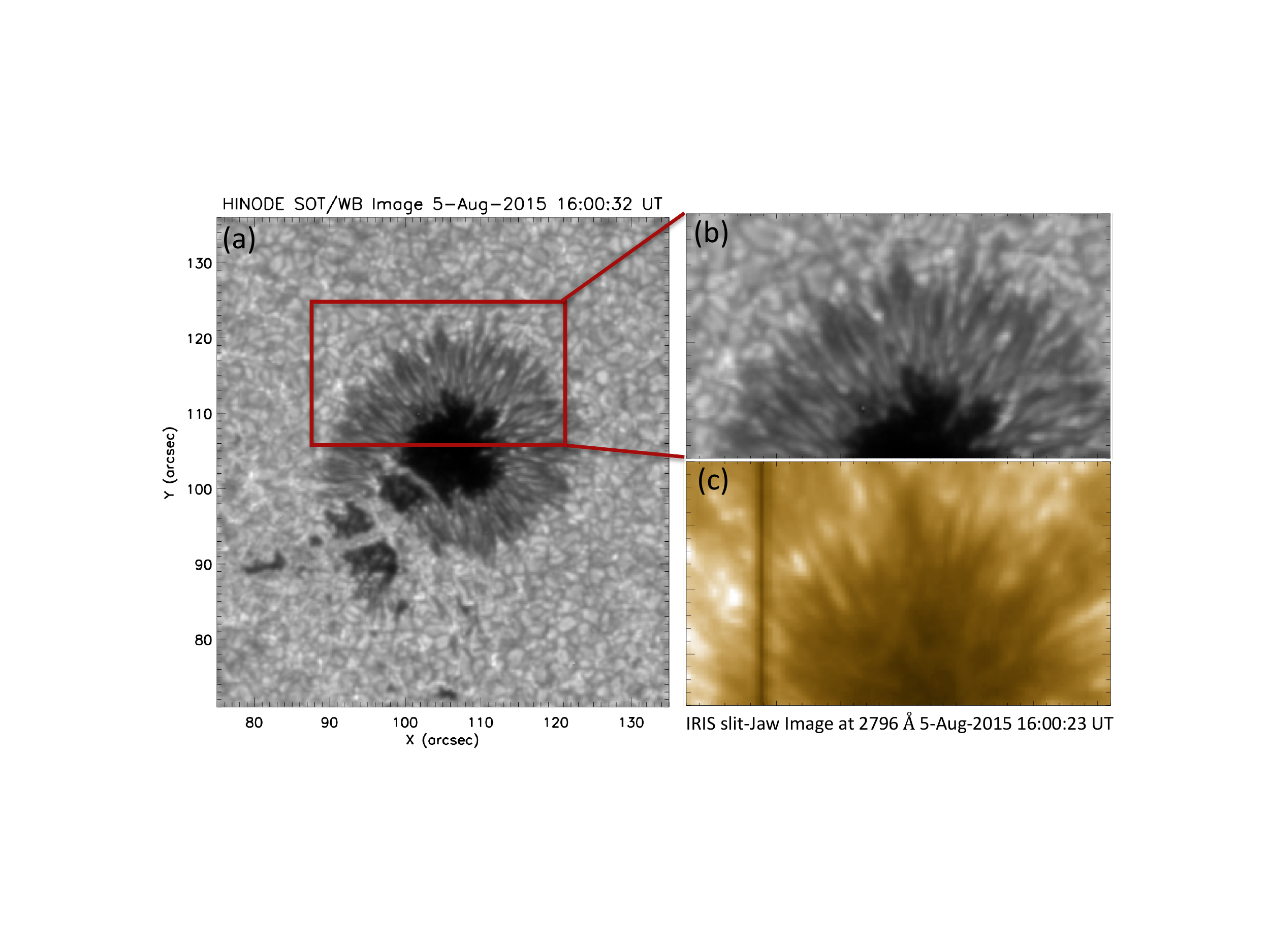}
	\caption{Context image of the sunspot in NOAA AR 12394, and the selected field of view (FOV) of interest in the penumbra. (a) G-band image from SOT/FG. (b) Zoomed in view of the selected FOV of the penumbra outlined by the red box in panel (a). (c) IRIS slit-jaw image at \MgII\ k 2796 \AA\ line, of the same time as in the image (a). A vertical shaded line marks the position of the slit at this particular time. }
	\label{fcontext1}
\end{figure*}

Both sunspots in our study were observed by Hinode (SOT): the first one by the filtergraph (FG) and the second one by the spectropolarimeter (SP). The FG data are from the magnetically sensitive \NaI\ 5896 line, and Stokes-V/I images are recorded. The SP repeatedly scanned about half of the penumbra of the second sunspot (in 6301.5 and 6302.5 \AA) in normal scan mode, thus, having full Stokes (I, Q, U, and V) data with a pixel size of 0.16 arcsec and a noise level of 10$^{-3}~I_c$ \citep{ichi08,lites13SP}. 

The FG G-band images have a pixel size of 0.22 arcsec, and the \NaI\ 5896 Stokes-V/I images have a pixel size of 0.32 arcsec. Because of a stationary bubble in the tunable filter there is some blurring in parts of the image, degrading the resolution and hindering detection of mixed-polarity field in Stokes-V/I images.

For the first sunspot (the leading part of NOAA AR 12394) on 05 August 2015, during 11:09:21--16:06:17 UT, IRIS ran a large coarse 8-step raster with a step cadence of 9 s, and a raster cadence of 73 s, thus resulting in a total of 245 rasters (OBS ID 3860109380). The exposure time for each slit position was 8 s. Slit-jaw images in two channels, \CII\ 1330 \AA\ and \MgII\ 2796 \AA, at every 18 s were recorded and have a pixel size of 0.33 arcsec. The IRIS slit width is 0.33 arcsec. The spectra have a slit step size of 2 arcsec and a pixel size of 0.33 arcsec in the vertical (Y-) direction, due to a binning in the Y-direction by a factor of two. 


%

The second sunspot penumbra (the leading part of NOAA AR 12680) was observed in September, 2017, from `2017-09-15T23:04:42.570' to `2017-09-16T03:52:25.510' (OBS ID 3624258055).  IRIS scanned medium dense 64-step rasters with the slit step size of 0.35 arcsec and the y-pixel size of 0.16 arcsec. The step cadence is 9 s and the raster cadence is 595 s, thus, resulting in 29 rasters in about five hours of observations. The exposure time for each slit position, also for this sunspot, is 8 s. The IRIS slit-jaw images in each of the four IRIS channels are recorded at a cadence of 37 s and have a pixel size of 0.16 arcsec. 

All IRIS data were already calibrated to level 2, i.e., dark-current subtracted, flat-fielded, corrected for geometrical distortion and wavelength calibrated. 

FG data were not available during this time of observation for the second sunspot, but SP observations are available, at a 7.5 minute cadence. SP scanned a part of the sunspot penumbra multiple times from `2017-09-15T22:48:06.937' to `2017-09-16T05:48:06.531'. To align Stokes-V scans with the IRIS \MgII\ images, we first align IRIS slit-jaw continuum intensity images of the \MgII\ k line in the far red wing (2832 \AA) (which provides high-contrast photospheric imaging) with the continuum intensity/G-band images obtained by SOT/SP/FG. The FG and SP data were calibrated using FG\_PREP and SP\_PREP routines, respectively, see \cite{lites13} for details.  

Due to their transient nature penumbral jets are often most clearly detected in running difference images. Therefore we use running difference slit-jaw images to calculate different properties of jets, listed in Tables \ref{t1} and \ref{t2}. However, depending on the timings of the image frames involved in creating running difference images, due to the limited slit-jaw image cadence, sometimes jets are not best seen in running difference images. In those cases we have used direct intensity \MgII\ slit-jaw images to estimate those numbers characterizing jets. 

We analyze the IRIS rasters for the \MgII\ k 2796 \AA\ line and create Dopplergrams by subtracting the intensities in the blueshifted and redshifted wings of the spectral line at fixed offset velocities from the rest wavelengths e.g., in our case for better visibility of both blueshifts and redshifts we choose $\pm$50 \kms\ \citep{depo14Science}. To reduce any local fluctuations in the Dopplergrams, we averaged Dopplergrams created by integrating the signal over a range of 10 \kms\ centered around $\pm$50 \kms.  The Dopplergrams for \MgII\ 2796 k line plasma provide structure and dynamics (redshift and blueshift) of chromospheric plasma.   

An overview image of each of the sunspot penumbrae is shown in Figures \ref{fcontext1} and \ref{fcontext2}.

\begin{figure*}
	\centering
	\includegraphics[trim=2.9cm 4cm 2cm 3.4cm,clip,width=0.94\textwidth]{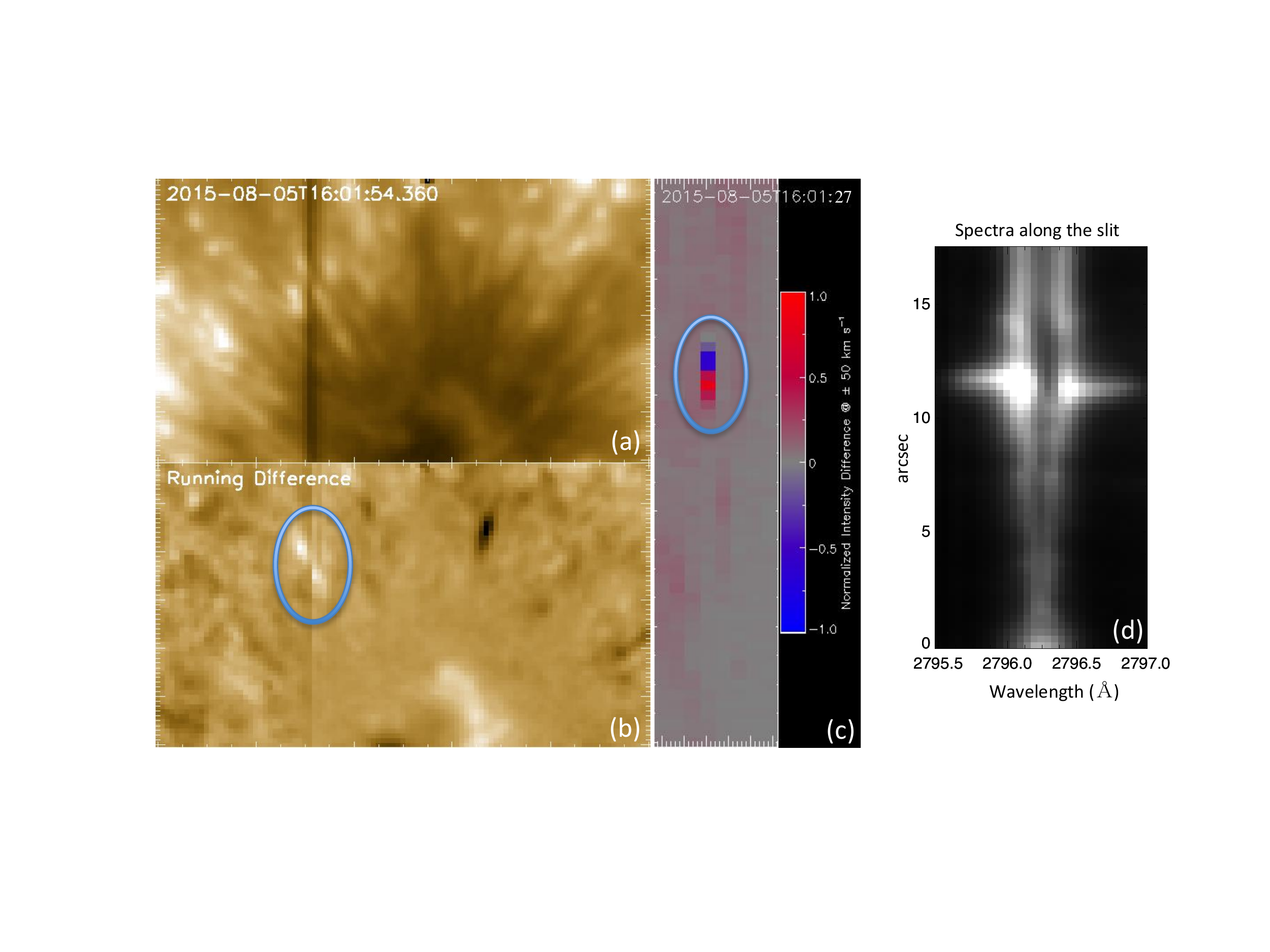}
	\caption{An example of a large penumbral jet from the sunspot shown in Figure \ref{fcontext1}. The IRIS slit was located across the middle of the jet during the jet's peak time. (a) \MgII\ 2796 \AA\ slit-jaw image. (b) Running difference image. The jet is outlined by an oval shaped blue circle. Note that a part of the jet is obscured by the slit. An animation for panels (a) and (b) is available, see ``SJ\_Pen1.mp4": each jet listed in Table \ref{t1} is marked in the movie by a red arrow during its peak-visibility time, with the arrow also displayed in the nearest earlier and nearest later frames to help viewer focus on the location of the jet. (c) The Dopplergram shows that the opposite sides of the jet have opposite (red/blue) Doppler shifts. Of the 8-slits of the 241st raster the fourth slit caught this event. The time noted on the Dopplergram image is of the first slit of the 8-step raster. An animation of panel (c) is available, see ``Doppler\_Pen1.mp4". Jets in the Dopplergram movie are marked by white arrows. Similar to that in the slit-jaw movie each jet location in the Dopplergram movie is marked by an arrow in three consecutive frames, the middle frame corresponding to the peak-visibility time. In panel (d) spectra along the slit are displayed.}
	\label{f2}
\end{figure*}

\begin{figure*}
	\centering
	\includegraphics[width=0.885\textwidth]{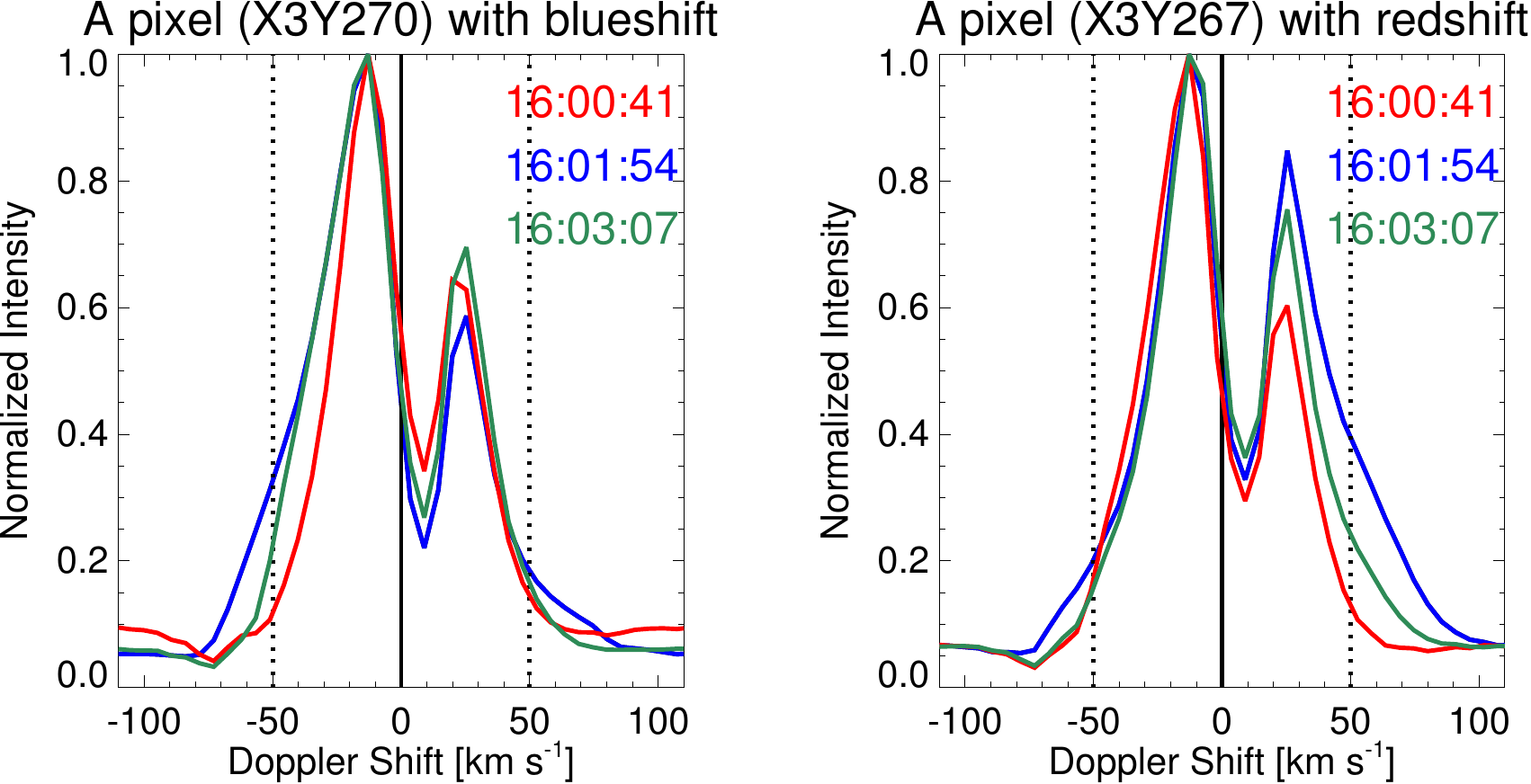}
	\caption{Line profiles of two pixels inside the jet region displayed in Figure \ref{f2}: one at a blueshifted location, and the other at a redshifted location. The red, blue, and green profiles are respectively for the times before, during and after the event. The vertical solid line marks zero Dopplershift. Two dashed vertical lines mark the locations of Dopplershifts at $\pm$50 \kms. In the left/right panel, clearly a blueshift/redshift at $\pm$50 \kms, in the wings of the blueshifted/redshifted pixel, can be seen at 16:01:54 UT.}
	\label{lp}
\end{figure*}

\section{Results} \label{resu}
We created Dopplergrams of IRIS spectral rasters of both sunspots for the \MgII\ k line, in which penumbral jets are best seen. 
As mentioned earlier, Dopplergrams are intensity differences at fixed offsets (here it is $\pm$50 \kms) in blue and red wings. We aim to detect any signatures of blueshift and redshift next to each other along the slit at the location where the slit cuts across the jet. If the field in the spire of large penumbral jets is twisting, there should be signatures of it in the spectral profiles and in Dopplergrams.

\subsection{Sunspot penumbra from the leading spot of NOAA AR 12394}
We show in Figure \ref{fcontext1} the full sunspot and the FOV selected for our investigation. Note that we selected only the northern part of the penumbra to avoid viewing the activity in the light bridge present in the southern part of the sunspot.   

In Figure \ref{f2}, we show an example large penumbral jet in a \MgII\ k line slit-jaw image, and a running difference image during the peak time of the jet. The respective Dopplergram and spectra are displayed in panels (c) and (d), respectively. Panels (a) and (b) are a frame from the movie SJ\_Pen1.mp4, and panel (c) is a frame from the movie Doppler\_Pen1.mp4. During the peak of the jet the IRIS slit crossed the middle of the jet at an angle of $\sim30\degree$ to the jet's long dimension. In the Dopplergram, there is a clear blueshift and redshift pattern next to each other, along the slit, when the slit was located in the middle of the jet during the jet's peak time. We interpret these flows as the signature of twisting motion of the magnetic field in the spire of the large penumbral jet. Consistent with the Dopplergram, the spectral profiles at this time show obvious shifts towards the blue/red wing of the line at respective pixels, shown in panel (d).

In Figure \ref{lp}, we plot spectral profiles for a blueshifted and a redshifted pixel during, before, and after the event, along the fourth slit of raster number 241. Consistent with Figure \ref{f2}, a clear blueshift and redshift in blue and red wings, respectively, are seen during the event time as compared to the profiles before and after the event. No significant shift is seen at the line center k3 during the peak event time with respect to the quieter times.  This might suggest that the k3 is formed in the background atmosphere and not inside of large penumbral jets. The same is also true for the k2v and k2r peaks. 

A dominant 6 -- 8 \kms\ redshift visible in the Figure \ref{lp} could be due to the inverse Evershed flow \citep{ever09a}. All over the penumbra such chromospheric downflows ranging 2 -- 10 \kms\ are present.

\begin{figure}
	\centering
	\includegraphics[trim=4.3cm 0cm 6.72cm 0.15cm,clip,width=\columnwidth]{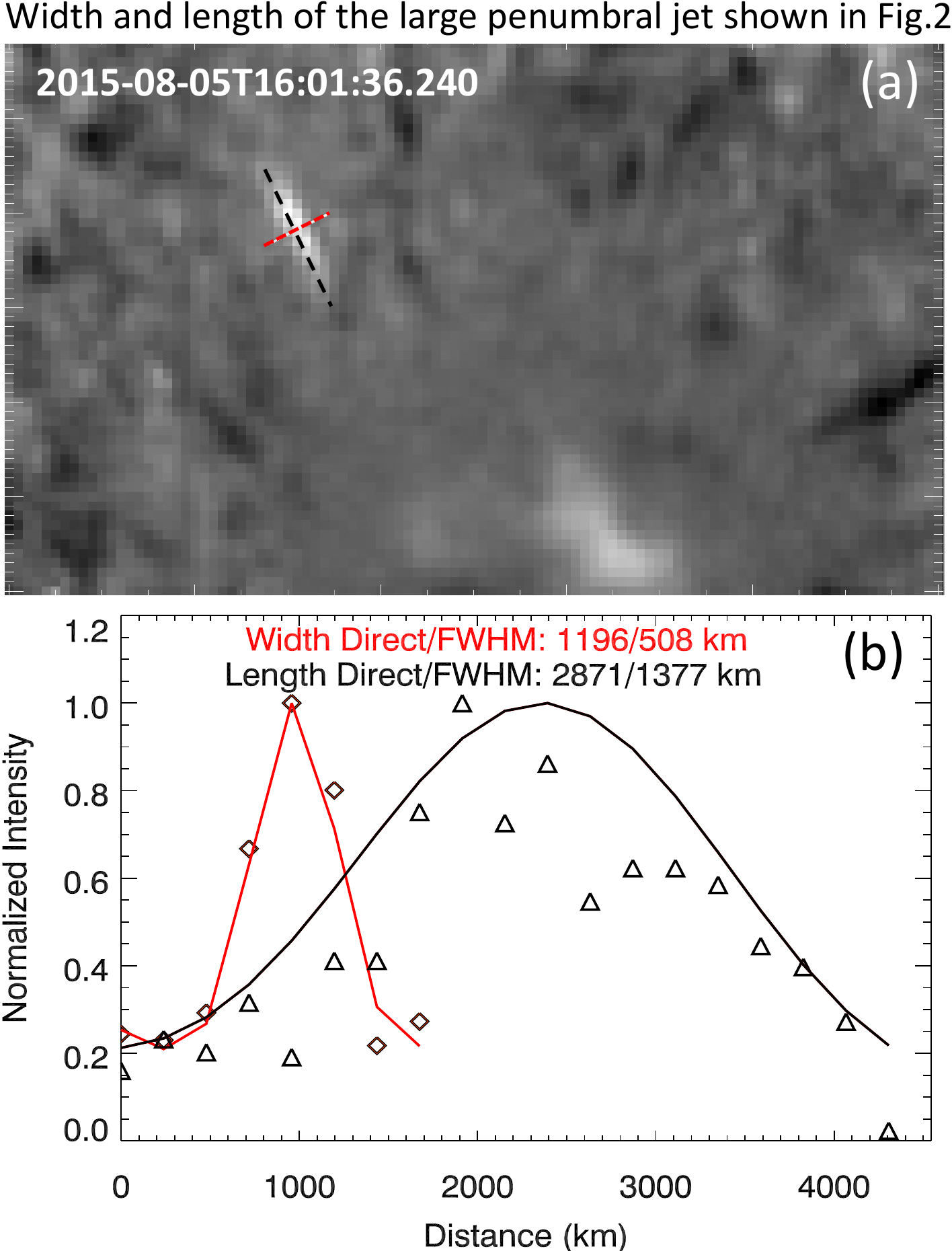}
	\caption{Width and length measurements of the example large penumbral jet shown in Figure \ref{f2}. (a) Running difference image with a red/black cut across width/length of the jet. The intensity plots along these cuts (diamonds -- width; triangles -- length), with their fitted Gaussian (solid lines), are shown in panel (b). The data fits the Gaussian acceptably well. Both the width and length directly estimated from the intensity enhancements with respect to the background (i.e., from the Gaussian width) and by FWHM of the Gaussian fitted function are printed, and are also listed in Table \ref{t1}. During the peak of the jet the IRIS slit was centered on the middle of the jet, partially obscuring the jet and thereby complicating the measurements. Thus, we measured the jet width and length in the nearest earlier frame of the IRIS data.}
	\label{width1}
\end{figure}

\begin{figure*}
	\centering
	\includegraphics[trim=4.6cm 7.22cm 3.9cm 4.7cm,clip,width=0.85\textwidth]{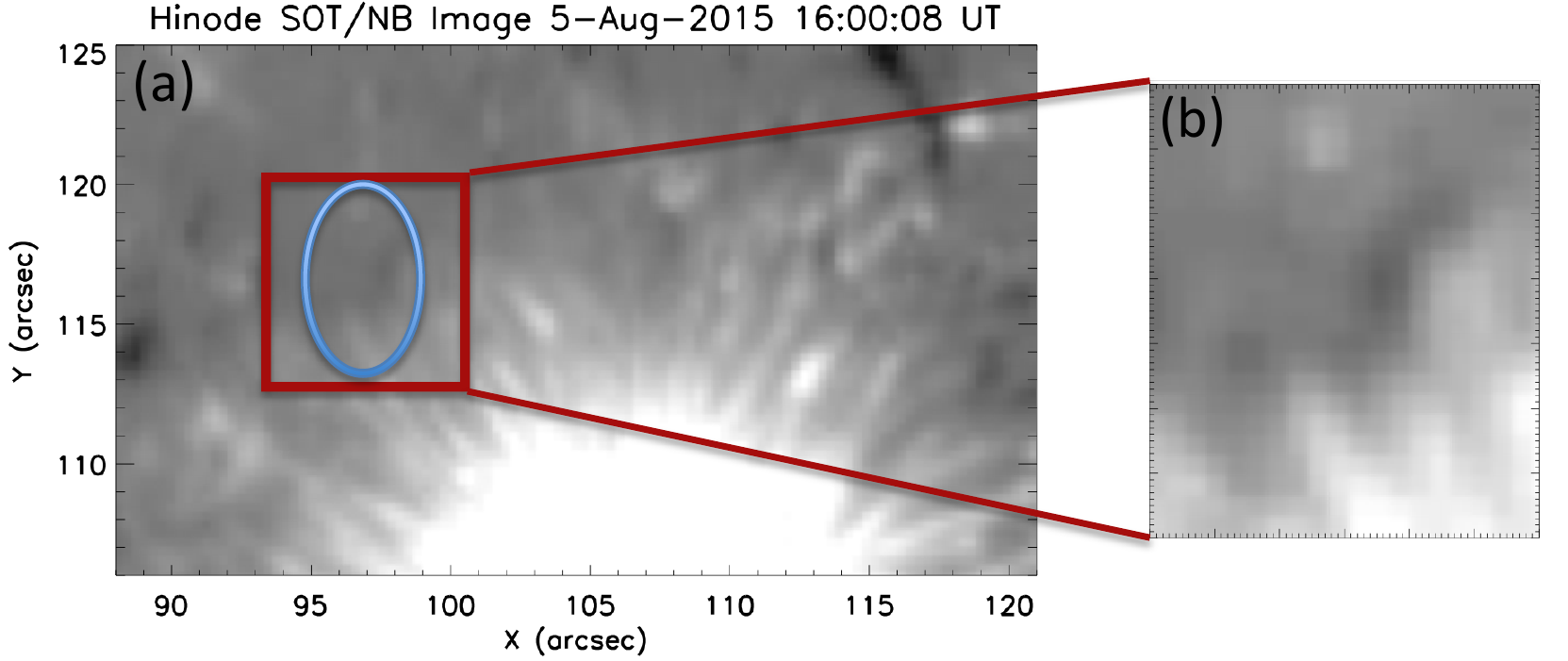}
	\caption{Narrow-band (NB) FG Stokes-V/I image, qualitatively equivalent to a magnetogram, in the \NaI\ 5896 \AA\ line near the time of the large penumbral jet shown in Figure \ref{f2}. The magnetic field is given in arbitrary units, ranging $\pm$1, i.e., the displayed field is symmetric about zero. The oval is centered on the jet-base location. In the right panel we enhance the intensity of the jet base to show the presence of mixed polarity field. Note that the magnetogram is blurred by a stationary bubble in the tunable filter.}
	\label{fgmg}
\end{figure*}

\begin{deluxetable*}{ccccccc}[hpt]
	\setlength{\tabcolsep}{4pt} 
	\tablenum{1}
	\tablecaption{List of six large penumbral jets caught by the IRIS slit in penumbra of the leading spot of NOAA AR 12394
		 \label{t1}}
	\tablewidth{0pt}
	\tablehead{\colhead{Peak-visibility time (UT)} & \colhead{Length (km)\tablenotemark{a}} & \colhead{Width (km)\tablenotemark{a} }  & \colhead{Lifetime\tablenotemark{b}} & \colhead{Speed\tablenotemark{c}} & \colhead{Dopplergram feature} & \colhead{Mixed-polarity}\\
		\colhead{SJ/Dopplergram movie} & \colhead{Direct/FWHM}  & \colhead{Direct/FWHM} & \colhead{(s)} & \colhead{(\kms)} &  \colhead{(with comments)} & \colhead{flux in FG} }
	\startdata
	12:32:14/12:31:46  &2871/1376 &1914/475 & 109 &26 & red, slit on head & no data \\ 
	12:52:32/12:52:22 &2632/993 &1196/451 &  91 & 29 & blue, slit on head & no data \\
	12:54:57/12:54:48 & 3349/1072&1675/455 & 54 & 62 & red, slit on tail& no data \\
	13:01:01/13:00:51  & 3589/1047 &1196/401 & 73 & 49 & blue, early phase & no data \\
	15:15:51/15:15:23  &  3828/1016 &1436/675 & 73 & 52 & red, last phase &  yes\tablenotemark{d} \\
	16:01:54/16:01:27 & 2871/1377 &1196/508 &  90 & 32 & blue, red (twisting) & yes \\ 
	\hline
	average & 3190/1147 &1436/494 & 82 & 42 & -- & --  \\
	\enddata 
	\tablenotetext{a}{Lengths and widths calculated from direct intensity enhancements with respect to their background might have an uncertainty of one to two pixels, i.e., up to 480 km.} 
	\tablenotetext{b}{Uncertainty in the estimated lifetimes is less than two frames, i.e., less than 37 s.}
	\tablenotetext{c}{Based on the estimated uncertainties for lengths and lifetimes of large penumbral jets, the uncertainty in the estimated speeds can be up to 13 \kms. The speeds are calculated from the length and the lifetime of each of the jets. We also verified these numbers from the speeds derived from time-distance measurements. Note that the given speed might well be different/lower from the actual speeds due to projection effects.}
	\tablenotetext{d}{FG Stokes-V image is blurred, similar to the one shown in Figure \ref{fgmg}.}
	
	\tablecomments{ The displayed time for a penumbral jet in the slit-jaw movie (SJ\_Pen1.mp4) and in the Dopplergram movie (Doppler\_Pen1.mp4) can be different, as evident from the first column in the table, because the time shown in the Dopplergram movie is the time of the 1st slit position for each raster. Width and length are measured at the peak-visibility time of the jet, which might or might not be the time of the slit position on the jet. Due to foreshortening the listed values of lengths and speeds of the jets are their lower limits.}
\end{deluxetable*}

In Figure \ref{fgmg}, we show an FG Stokes-V image in \NaI\ 5869 \AA\ line of the same FOV as IRIS. Note that the magnetogram is blurred by a stationary bubble in the tunable filter. Therefore in the right panel of the Figure we augment the small FOV of interest that covers the base of the large penumbral jet displayed in Figure \ref{f2}. The presence of mixed-polarity field is obvious at the base of the example large penumbral jet. This location is also apparently where the tails of several penumbral filaments converge.

Because of the coarseness of the IRIS rasters (2 arcsec steps) the probability of catching a penumbral jet with the IRIS slit is small. Nonetheless, over the period of our observations we did catch five additional large penumbral jets by the IRIS slit in this sunspot penumbra. These jets displayed clear signature in Dopplergrams and are listed in Table \ref{t1}. 

Each jet listed in Table \ref{t1} is marked by a red arrow in the slit-jaw movie (SJ\_Pen1.mp4) and by a white arrow in the Dopplergram movie (Doppler\_Pen1.mp4) during their peak-visibility time. To help viewers focus on the locations of large penumbral jets in the slit-jaw and Dopplergram movies, we have also added an arrow for each large penumbral jet in the frames just before and just after the peak-visibility time. The peak-visibility times of these jets, with their estimated lengths, widths (e.g., Figure \ref{width1}), lifetimes (calculated from the movie SJ\_Pen1.mp4), speeds (calculated from their lengths and lifetimes, also verified with time-distance method) and whether they have blueshift/redshift in Dopplergrams and have mixed-polarity field at their base magnetograms, are also listed in the Table \ref{t1}. FG data was available only for the last two jets, listed in Table \ref{t1}.

Due to the raster scan being coarse and jets being transient, often only a part of a jet (e.g., only head or only tail) is caught by the IRIS slit. Moreover, the slit often did not cross the jets during their peak times. These factors might be why some jets show only blueshift or only redshift. Some of the large penumbral jets reoccur at the same locations for several minutes, thus, share some properties of Ellerman bombs \citep{elle17,rutt13}. However, Ellerman bombs are mostly found outside sunspots (not inside the sunspot penumbra); see Section \ref{diss} for a discussion on the similarities and differences between Ellerman bombs and large penumbral jets. Some jets start in the penumbra extending well outside it and are caught at the outer periphery by the slit. A few of the strong events do not have any signatures in the Dopplergrams. We have not included those jets in the table.

\subsection{Sunspot penumbra from the leading spot of NOAA AR 12680}

\begin{figure*}
	\centering
	\includegraphics[trim=0.6cm 4.4cm 1.1cm 3.3cm,clip,width=\textwidth]{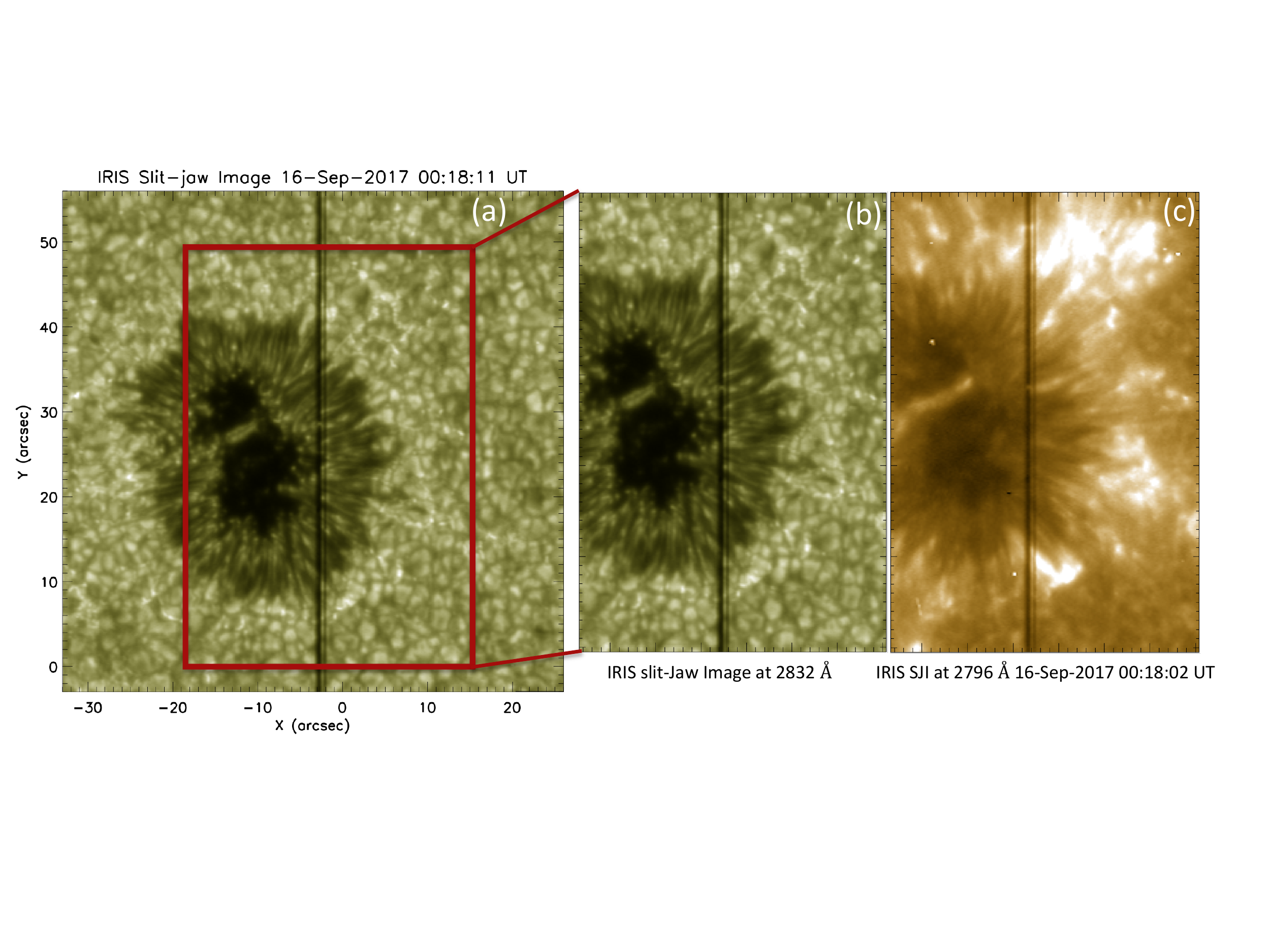}
	\caption{Context image of the sunspot in NOAA AR 12680, and the selected field of view (FOV) of interest of the sunspot penumbra. (a) IRIS slit-jaw \MgII\ continuum image at 2832 \AA. (b) Selected FOV showing the close-up view of the penumbra. (c) IRIS slit-jaw image at \MgII\ 2796 \AA\ line, of the same time as in (a). The shaded vertical line in each panel marks the position of the slit at the time of the observation. The sunspot is almost at the solar disk center.}
	\label{fcontext2}
\end{figure*}

\begin{figure*}
	\centering
	\includegraphics[trim=4cm 0.65cm 8cm 0cm,clip,width=0.92\textwidth]{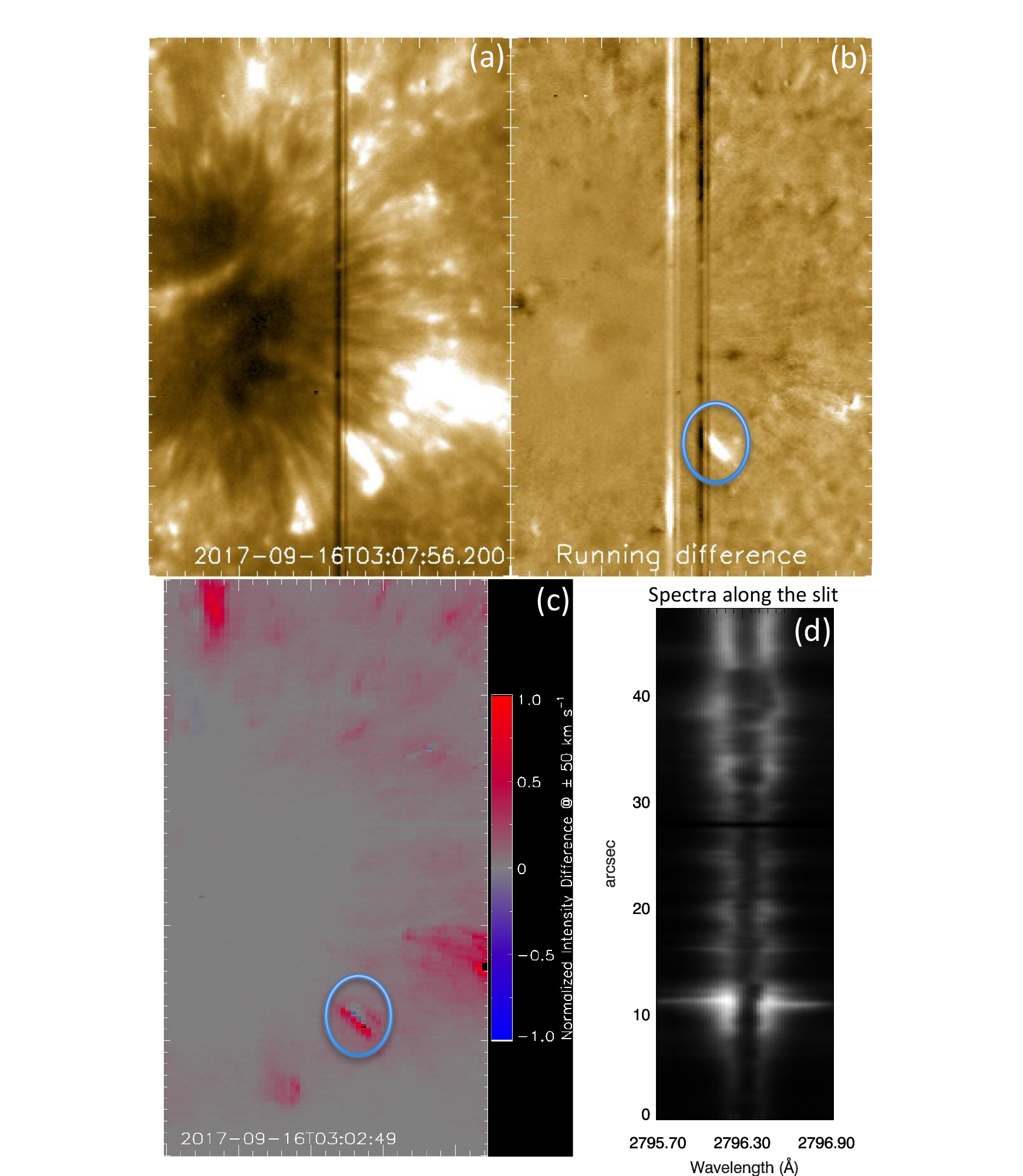}
	\caption{An example large penumbral jet from the sunspot shown in Figure \ref{fcontext2}. The IRIS slit covered the jet in multiple steps. (a) \MgII\ 2796 \AA\ slit-jaw image. (b) Running difference image with a blue oval centered on and encircling the example jet. (c) The Dopplergram showing that the opposite sides of the jet have opposite (red/blue) Doppler shifts. Note that a part (tail) of the jet is obscured by the slit in (a) and (b). The position of 38th slit in the 24th raster is shown as a dark shaded line in (a) and (b). The time noted on the Dopplergram image is of the first slit of the 64-step raster. In (d) \MgII\ spectra corresponding to the slit in the middle of the jet are displayed. Animations for panels (a), (b): ``SJ\_Pen2.mp4", and (c): ``Doppler\_Pen2.mp4" are available. Each jet listed in Table \ref{t2} is marked by a red arrow in the slit-jaw movie (SJ\_Pen2.mp4) and by a white arrow in the Dopplergram movie (Doppler\_Pen2.mp4) during its peak-visibility time, with the arrow also displayed in the nearest earlier and nearest later frames of the movies to help viewer focus on the location of the large penumbral jet. } 
	\label{f6}
\end{figure*}

In the same way as for the first sunspot we created Dopplergrams for this sunspot penumbra. A context image of the sunspot is shown in Figure \ref{fcontext2}. The IRIS raster for this sunspot penumbra is 64-step, a medium dense raster, thus, although the raster cadence is worse, the dense coverage leads to better spatial coverage along penumbral jets. In Figure \ref{f6}, we display one example large penumbral jet, see panels (a) and (b) (and movie: SJ\_Pen2.mp4). In panel (c) we plot the Dopplergram (see movie: Doppler\_Pen2.mp4), and the location of the jet is outlined with an oval. Blueshifted and redshifted pixels next to each other across the slit can be clearly noticed at the jet location. Because the slit was on the jet for several consecutive steps a clear extension of the blueshift/redshift along the jet can be noticed.

In panel (d) of Figure \ref{f6}, the spectra along the slit at one of these times (when the slit crossed the jet) are plotted. Blueshift and redshift (line-wing broadening in the blueshifted and redshifted pixels) are consistent with the blueshift/redshift observed in the Dopplergram in panel (c). We again interpret the blueshift and redshift present next to each other across the large penumbral jet as twisting/untwisting of the magnetic field in the spire of the jet.

\begin{figure*}[h]
	\centering
	\includegraphics[width=0.885\textwidth]{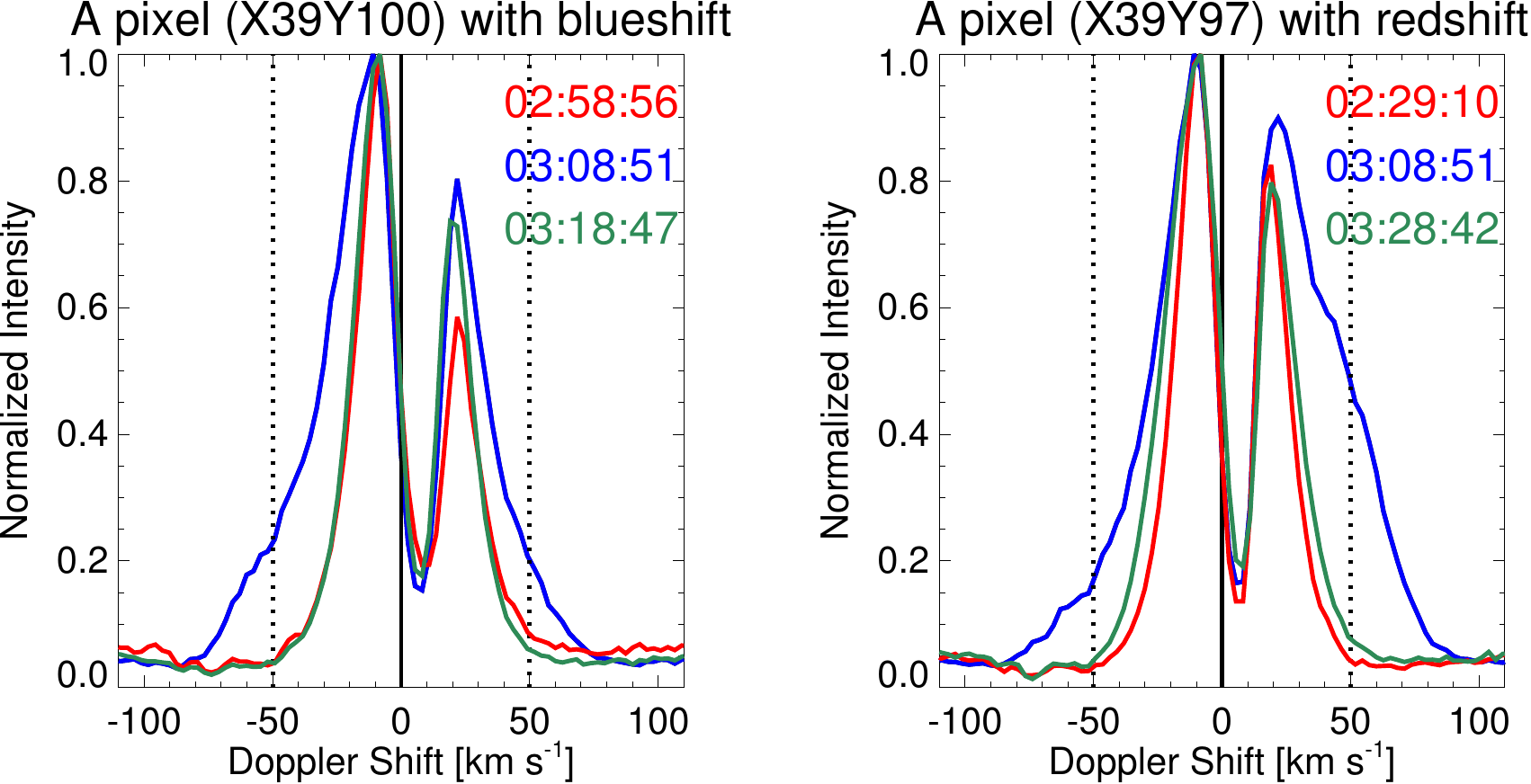}
	\caption{Line profiles of two pixels inside the example large penumbral jet region shown in Figure \ref{f6}: one at a blueshifted location, and the other at a redshifted location, both along the 38th slit of the raster number 24 (each raster has 64-steps). The red, blue and green profiles are respectively for times before, during and after the event. The vertical solid line marks zero Dopplershift. Two dashed vertical lines mark the locations of Dopplershifts at $\pm$50 \kms. In the line wings of the left/right panel at $\pm$50 \kms\ a clear blueshift/redshift can be seen at 03:08:51 UT.}
	\label{lp2}
\end{figure*}

In Figure \ref{lp2}, we display the line profiles of two pixels across the slit -- one blueshifted, one redshifted. There are obvious blueshift and redshift present in the respective line profile wings. It is important to note that although there is excessive blueshift and redshift in the pixels of corresponding profiles, there is a strong broadening present in the line during the event. Possible explanations for the strong line broadening are unresolved motions, e.g., from turbulence, or increased opacity because of heating (due to increased temperature and density). In either case, the excessive shifts in the blue and red wings indicate that there is twisting motion of the magnetic field lines that channel the jet plasma.   

\begin{figure}
	\centering
	\includegraphics[trim=7.8cm 5.1cm 9.4cm 2.6cm,clip,width=\columnwidth]{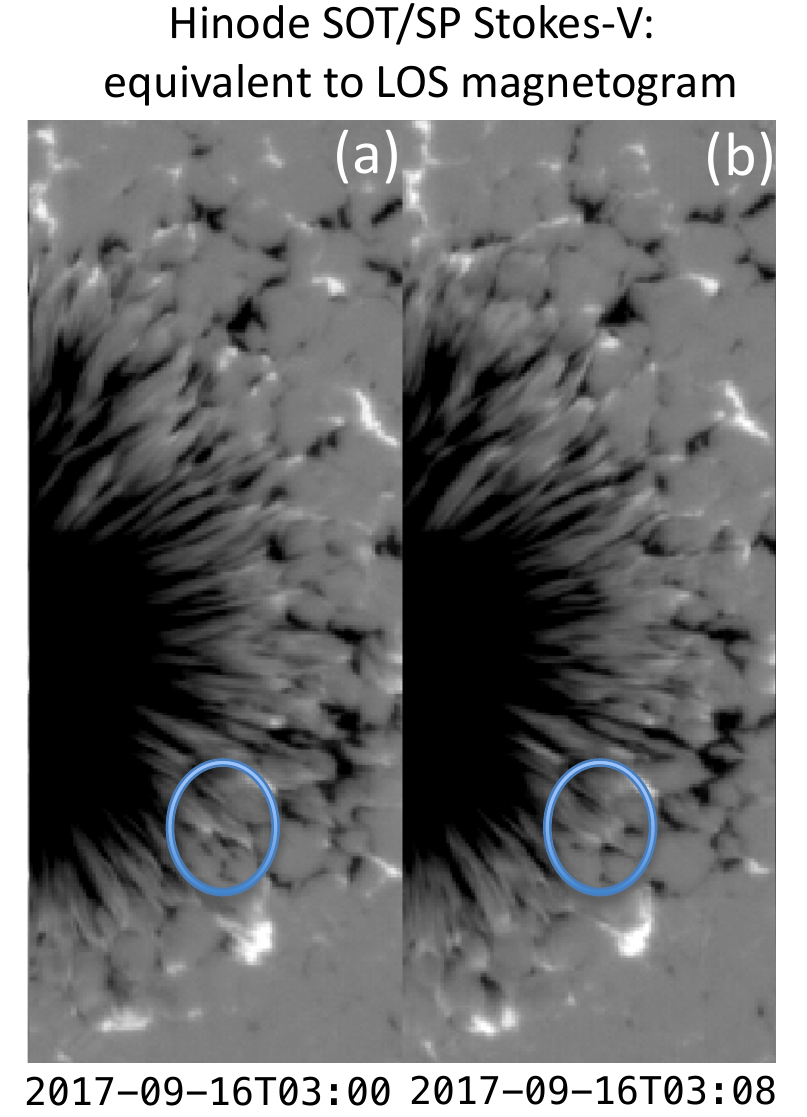}
	\caption{SP Stokes-V images (qualitatively equivalent to magnetograms) displaying the base of the large penumbral jet shown in Figure \ref{f6}. The magnetic field is given in arbitrary units, ranging $\pm$1, i.e, the displayed field is symmetric about zero. The blue oval shape is centered on the base of the jet. The two panels (a) and (b) show Stokes-V scans starting before and just after the jet event. The time given on the bottom of each panel belongs to the first spectral slit of the SP scan of the given FOV of the sunspot penumbra. An animation (SP\_StokesV.mp4) is available with similar two frames for each jet listed in Table \ref{t2}.}
	\label{fgmg2}
\end{figure}

We looked at the base of this jet in Stokes-V maps from the SP observations. Two image frames of the sunspot penumbra scanned before and after the jet are plotted in Figure \ref{fgmg2}. Clear locations of positive (minority) polarity flux can be noticed just before the jet event, and the field disappears after the jet is over. This is consistent with flux cancellation being involved in the jet production as suggested by \cite{tiw16} for large penumbral jets, and by \cite{pane17,pane18} for EUV and X-ray coronal jets. 

It is important to note that the example jet shown in Figure \ref{f6} is a coupled jet (there are two jets occurring closely enough to appear as one). The tails of two penumbral filaments (with positive polarity magnetic field) involved in the jets are seen inside the oval outlined in the Figure \ref{fgmg2} of Stokes-V images.

A dominant overall general downflow ranging 2 to 10 \kms\ is found also in this sunspot penumbra, which again can be attributed to the chromospheric inverse Evershed flow \citep{ever09a}. 

\begin{figure*}
	\centering
	\includegraphics[trim=1.5cm 6cm .5cm 0.6cm,clip,width=\textwidth]{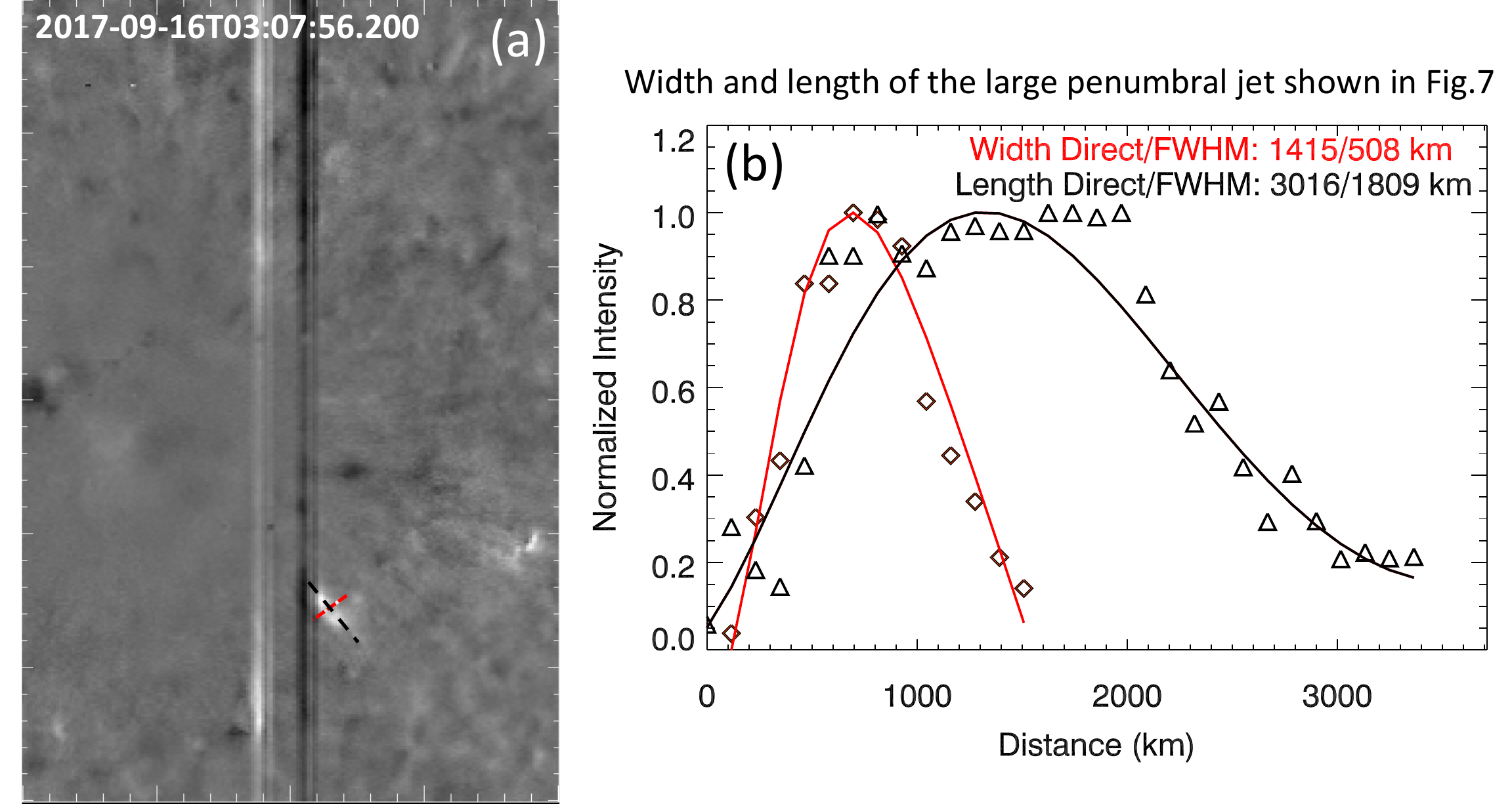}
	\caption{Width and length measurements of the example large penumbral jet shown in Figure \ref{f6}. (a) Running difference image with a red/black cut across width/length of the jet. The intensity plots along these cuts (diamonds -- width; triangles -- length), with their fitted Gaussian (solid lines), are shown in panel (b). Again, the data fits the Gaussian acceptably well. Both the width/length directly estimated from the intensity enhancement with respect to the background (i.e., from the Gaussian width) and by FWHM of the Gaussian fitted function are printed, and are also listed in Table \ref{t2}.}
	\label{width2}
\end{figure*}

\begin{deluxetable*}{ccccccc}
	\setlength{\tabcolsep}{1pt} 
	\tablenum{2}
	\tablecaption{List of 11 large penumbral jets caught by the IRIS slit in penumbra of the leading spot of NOAA AR 12680 \label{t2}}
	\tablewidth{0pt}
	\tablehead{	\colhead{Peak-visibility time (UT)} & \colhead{Length (km)\tablenotemark{a}} & \colhead{Width (km)\tablenotemark{a}} & \colhead{Lifetime\tablenotemark{b}} & \colhead{Speed\tablenotemark{c}} & \colhead{Dopplergram feature} & \colhead{Mixed-polarity flux in SP}\\
		\colhead{SJI/Dopplergram movie} & \colhead{Direct/FWHM} & \colhead{Direct/FWHM} & \colhead{(s)}& \colhead{(\kms)} & \colhead{(with comments)} & \colhead{(cancellation rate\tablenotemark{f}: Mx h$^{-1}$)}}
	\startdata
	23:58:48/23:54:19\tablenotemark{n}  &1856/853 & 1392/346 & 112 & 17 & dominant blue, red (twisting)\tablenotemark{d} & yes ($\sim$2$\times10^{18}$)\\ 
	23:59:26/23:54:19\tablenotemark{s}   &2784/740& 1392/607 & 73 & 38 &blue on head, red (twisting)& yes ($\sim$8$\times10^{17}$)\\ 
	00:18:39/00:14:09 & 2900/1488 & 1392/437 &  73 & 40 & blue, red (twisting) & yes ($\sim$1$\times10^{18}$) \\
	00:59:34/00:53:50& 5336/3961 & 1740/490         & 112 & 48 & red blue (twisting) & yes\tablenotemark{e} \\  
	01:10:07/01:03:46& 5568/1041 & 1392/501         & 149 & 38 & red\tablenotemark{d}  & yes \\  
	02:18:19/02:13:12 & 3364/2660 & 1160/431 & 186 & 18 & blue, red (twisting) &  yes \\
	02:26:23/02:23:08 & 2900/788 & 928/360 & 112 & 26 &dominant blue, red (twisting)\tablenotemark{d}  &  yes \\
	03:08:33/03:02:49 &3016/1809& 1415/508 & 110 & 28 &blue, red (twisting) & yes ($\sim$9$\times10^{17}$)  \\ 
	03:18:28/03:12:44 &1856/919 & 1160/275 & 72 & 26 &blue on head, red (twisting) & yes \\ 
	03:29:38/03:22:39 &2552/738 & 1160/415  & 112 & 23 &red, blue (twisting) & yes \\ 
	03:47:37/03:42:30 &2320/566& 1044/355 & 111 & 21&red, blue (twisting) & yes \\ 
	\hline
	average & 2826/1415 &1289/430 &  111 & 30 & -- & yes ($\sim$1.2$\times10^{18}$) \\ 
	\enddata
	\tablenotetext{a}{Lengths and widths calculated from direct intensity enhancements might have an uncertainty of one to two pixels, i.e., up to 240 km.}
	\tablenotetext{b}{Uncertainty in the estimated lifetimes is less than two frames, i.e., less than 75 s.}
	\tablenotetext{c}{Based on the estimated uncertainties for lengths and lifetimes of large penumbral jets, uncertainty in the estimated speeds can be up to 4 \kms. Similar to that in Table \ref{t1}, the speeds are calculated from the length and the lifetime of each of the jets. We also verified these numbers from the speeds derived from time-distance measurements. The given speeds might be different/lower from the actual speeds due to projection effects.}
	\tablenotetext{n}{in northern part of penumbra.}
	\tablenotetext{s}{in southern part of penumbra.}
	\tablenotetext{d}{twisting is ambiguous in this case.}
	\tablenotetext{e}{a few SP slits are missing at the jet base.}
	\tablenotetext{f}{The uncertainty in the estimated cancellation rates is on the order of a factor of two to three.}
	
	\tablecomments{The displayed time for a penumbral jet in the slit-jaw movie (SJ\_Pen2.mp4) and in the Dopplergram movie (Doppler\_Pen2.mp4) can be different, as evident from the first column in the table, because the time shown in the Dopplergram movie is the time of the 1st slit position for each raster. Due to foreshortening the listed values of lengths and speeds of the jets are their lower limits. In a few cases when positive polarity magnetic flux (at the tails of penumbral filaments) can be clearly separated and flux cancellation is clearly visible, flux cancellation rates are estimated from $B_z$ derived from the ME inversions of the SP data at CSAC/HAO.}
\end{deluxetable*}

Ten other large penumbral jets in this sunspot penumbra, with their different properties i.e., lengths, widths (see, Figure \ref{width2}), lifetimes (calculated from the movie SJ\_Pen2.mp4), speeds (calculated from lengths and lifetimes, also verified with time-distance method), Dopplergram features, and mixed-polarity field in their SP Stokes-V images, are listed in Table \ref{t2}. These jets are also marked by red and white arrows in the movies SJ\_Pen2.mp4 and Doppler\_Pen2.mp4, respectively. Similar to the slit-jaw and Dopplergram movies in the first sunspot penumbra, also for the second sunspot penumbra (in the movies SJ\_Pen2.mp4 and Doppler\_Pen2.mp4) we point to each large penumbral jet listed in Table \ref{t2} by arrows in three frames: one during the peak-visibility time (given in Table \ref{t2}), and one each in the frames just before and just after the peak-visibility time.

Evidently, there are several jets displaying only mostly blueshift or only mostly redshift (e.g., the one at 01:03:46 UT in the movie Doppler\_Pen2.mp4, listed in Table \ref{t2}). Among other factors, the dominant flows of one sign could be because of one sign of the shift dominating over the other, the location of the slit (it covering only red or only blueshifted part of a jet; it not being perpendicular to the jet), or the jets themselves having no twisting/untwisting motions. One explanation for the dominance of one Doppler sign over the other, even in the presence of twisting, is that there are several different types of motions including field-aligned flows, which can offset the Dopplergrams to one color \citep{depo12,depo14Science}. 

Although the running difference images show the jets better due to jet's transient nature, these do not always enhance the jets due to the limited cadence of our data.  Therefore a few times while we see penumbral jets clearly in \MgII\ slit-jaw images, we do not see those jets in the running difference images so clearly. In those cases \MgII\ slit-jaw intensity images, rather than their running difference images, are used for estimating the relevant numbers listed in Table \ref{t2}. 

We also found several similar large jets forming just outside the sunspot penumbra, at the locations of obvious mixed polarity magnetic flux -- moving magnetic bipolar features. We found evidence of twist in Dopplergrams in these jets as well. These jets could be considered closer to (large) AR coronal jets \citep[\eg][]{ster16}, although these jets at the periphery of sunspots are still at much smaller scales than AR coronal jets. A detailed investigation of these jets emanating from moving magnetic features will be performed in a separate research work.

In a movie SP\_StokesV.mp4, we have marked by white circles the locations of each large penumbral jet listed in Table \ref{t2}. Note that the Stokes-V movie has only eleven frames, one for each example large penumbral jet. Each frame contains two maps of the penumbra, one scanned slightly before and the other slightly after the respective jet's peak time.

From the Milne-Eddington (ME) inversions of SP maps of the second sunspot, we could calculate flux cancellation rate in four of our large penumbral jets, which turns out to be, on average, of the order of 10$^{18}$ Mx h$^{-1}$. Note that it is not possible to always isolate the desired magnetic flux polarity from the complex surrounding in the sunspot penumbra for calculating the rate of flux cancellation involved in a penumbral jet formation. 
Further, sometimes the amount of magnetic field is too small to be detected at given pixels due to the noise and other artifacts in the non-converging ME inversions. Finally, due to the limited cadence of our SP data (7.5 minutes) the flux cancellation rate can not be accurately estimated.

\section{Discussion} \label{diss}

We investigated two sunspot penumbrae, each observed simultaneously by IRIS and Hinode, to find out whether large penumbral jets have twisting motions, and whether the jets always have mixed-polarity magnetic field at their base. Note that the selected sunspots were close to the solar disk center and therefore the visibility of the jets is not optimal, due to foreshortening. But the magnetograms do not have much projection effect for that very reason \citep[e.g.,][]{falc16}.  

We find for large penumbral jets an average length of 3000 km, an average width of 1350 km, an average lifetime of 95 s, and an average speed of 35 \kms. Some of these numbers fall on the longest end of the range provided by \cite{drew18}, who did not differentiate between microjets and large penumbral jets. The speeds of large penumbral jets in our sample are much smaller than that obtained for different penumbral jets by \cite{kats07} and \cite{tiw16}. This is likely due to the foreshortening (projection effect) due to our sunspots being located on the disk center. A part of the difference might also be due to the fact that different chromospheric lines were used in these studies (\CaII\ H line is used in the most earlier studies, whereas \MgII\ k line is used here). 

For some time now it has been observed that many QR, CH and AR coronal jets show twisting/untwisting motions, and that they have mixed-polarity magnetic field at their base. That mixed-polarity magnetic flux often progressively cancels before and through the time of coronal jet-onset, with recently reported  flux cancellation rate of 10$^{18}$ Mx h$^{-1}$ for CHs and QR jets \citep{pane16qr,pane17,pane18,pane18a}. AR jets have higher flux cancellation rate by an order of magnitude, however, with increased uncertainty in its measurement \citep{ster17}. We made a crude estimate of the flux cancellation rate in a few of our large penumbral jets, and find, consistently, the values of the order of 10$^{18}$ Mx h$^{-1}$. 

It must be noted that, for each jet, we have calculated flux cancellation rates based on two SP scans of the penumbra, one starting slightly before the jet and the other starting slightly after the jet, separated by 7.5 minute. Unless a filament is decaying, the downflows at the tails keep enhancing magnetic field there. 
Even if there is a decrease in the amount of magnetic field at the tails of penumbral filaments during a large penumbral jet, exact flux content involved in driving the jet is extremely difficult to accurately estimate. Moreover, because of the complex magnetic topology of the penumbra with several small-scale magnetic structures present in it \citep[see, e.g.,][and reference therein]{tiw17sunspot}, the observed Stokes profiles are often quite asymmetric, therefore ME inversion does not necessarily provide the most reliable results.  Further, the cadence of the observations is not good enough to attribute the measured flux cancellation exclusively to large penumbral jets. Clearly a faster cadence of spectropolarimetric observations with more sophisticated inversions \citep[see, e.g.,][]{del16} are needed to access accurate flux cancellation rates for penumbral jets. 

By using \MgII\ spectra for the two sunspot penumbrae in our study, observed with IRIS, we find that several large penumbral jets display blueshift and redshift, perpendicular to the jet direction, next to each other along the IRIS slit. This suggests that the magnetic field in the spire of large penumbral jets is twisting/untwisting. Because large penumbral jets have mixed-polarity field \citep[see][and this study]{tiw16}, the finding that they twist suggests that large penumbral jets might form the same way as QR, CH and AR coronal jets do. Note that, alternatively, propagating Alf\'ven waves that are generated in some other way might produce untwisting in the jet-spire of large penumbral jets. 

Penumbral microjets are proposed to form in a similar way as large penumbral jets, by reconnection in the lower atmosphere (higher photosphere) \citep{tiw16}. However, due to their small sizes and transient nature (short lifetime), a one to one correspondence between (chromospheric) microjets and magnetic features in the photosphere has not been established \citep{jurc08,tiw16}. Nonetheless, in a recent study, \cite{este18} could relate a number of penumbral microjets with the photospheric mixed-polarity field in the sides of penumbral filaments. Whether microjets are too narrow to display signatures of twisting with the current instrumentation (e.g., whether the SST data, which have higher spatial resolution than IRIS, show twisting motions in microjets) should be investigated. Therefore, whether all jets in sunspot penumbrae show twisting motions and whether all of them form in the same way remains to be seen.
 
\begin{figure*}[h]
	\centering
	\includegraphics[trim=0cm 0cm 0cm 0cm,clip,width=0.67\textwidth]{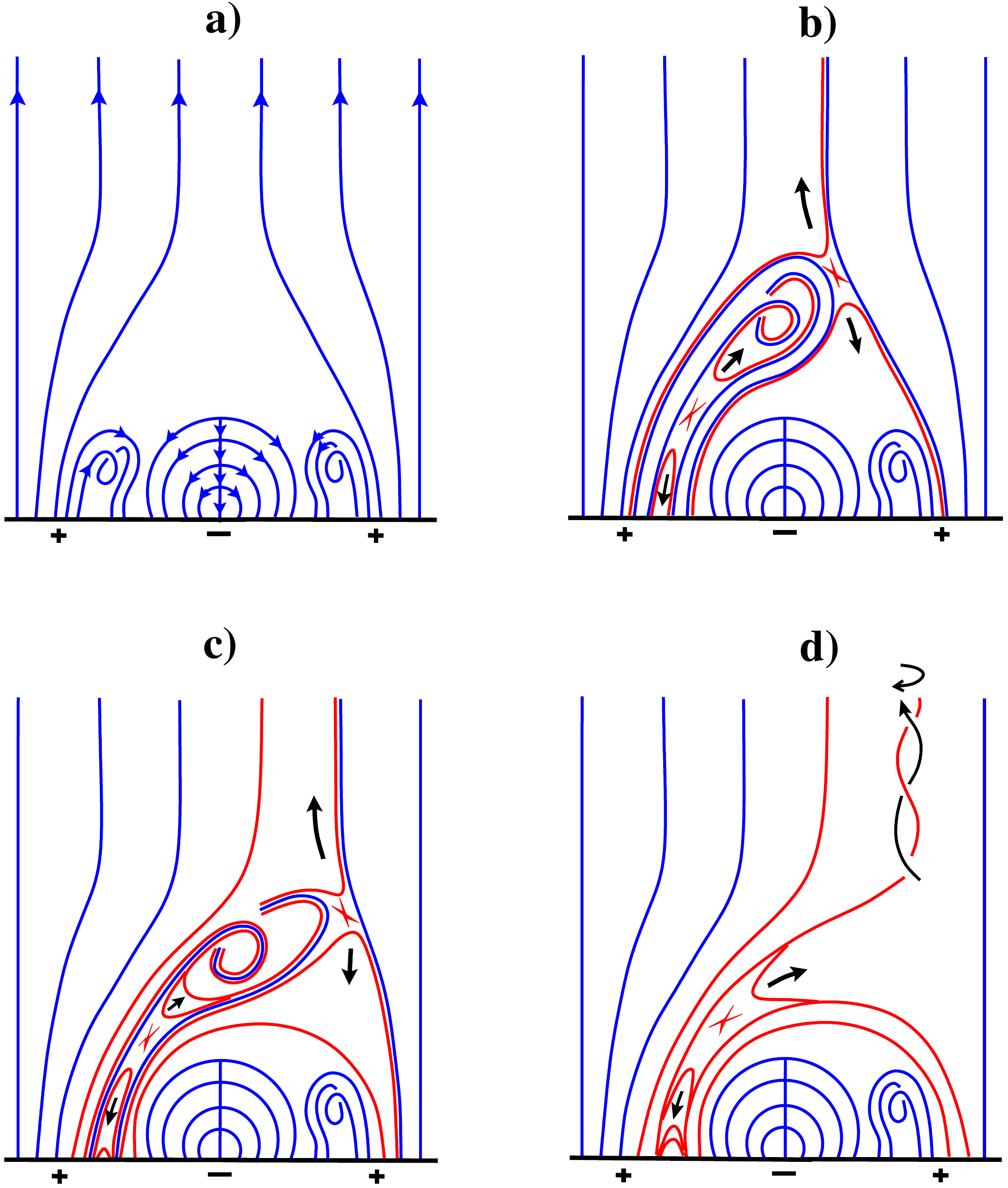}
	\caption{A schematic sketch depicting how penumbral jets could form, if they form in the same way as other coronal jets. Coronal jets (in CHs, QRs and ARs) untwist resulting from the eruption of a magnetic arcade whose core is greatly sheared and twisted and becomes a growing twisted flux rope as it erupts. Each drawing sketches the magnetic field in -- or the field's projection onto -- a slanted cross-cut plane through the tail of a penumbral filament in which the field direction is downward into the plane. The cross-cut plane slants away from the viewer, away from the head of the penumbral filament, so that lines of spine field surrounding the penumbral filament lie in the plane. The thick black line at the base of each drawing is some level low in the penumbral photosphere. The plus and minus signs give the polarity of the magnetic flux through that level. Blue lines are field lines that have not yet undergone reconnection in the eruption; red lines are field lines that have undergone reconnection in the eruption. Drawing (a) shows the direction and form of the field prior to the jet-eruption onset. The spine field has positive polarity and the field in the tail of the penumbral filament has negative polarity. Prior, convectively slowly-driven, evolutionary reconnection has built a sheared-core arcade over the polarity inversion line on each side of the tail, between the tail field and spine field. Due to the direction of the shear of the tail field relative to the surrounding spine field, the evolutionary reconnection has given the sheared-core arcade on the tail's left side right-handed shear and twist, and has given the arcade on the right side left-handed shear and twist. In the time between drawings (a) and (b), convection-driven flux-cancellation tether-cutting reconnection at the polarity inversion line of the left-side sheared-core arcade triggers the eruption of that arcade. Drawing (b) shows the envelop of the erupting arcade undergoing interchange reconnection with the encountered spine field, and shows the legs of the erupting arcade undergoing internal tether-cutting reconnection. Red Xs mark reconnection sites and black arrows represent magnetically-driven plasma outflows from the reconnection sites. Drawing (c) shows the interchange reconnection eating into the twisted-flux-rope core of the erupting arcade to open that twisted closed field. Drawing (d) shows the twist escaping and untwisting out along the newly opened field in the jet. Because the twist given to the open field is right-handed in this case, the spin direction of the untwisting in this large penumbral jet is left-handed, clockwise viewed facing the jet outflow direction.}
	\label{sketch}
\end{figure*}

If large penumbral jets form the same way as coronal jets in QRs, CHs and ARs, then these could form in the way described by the sketch in Figure \ref{sketch}. Note that the sketch describes left-handed untwisting of a large penumbral jet. A similar setup with the internal reconnection starting on the other side (compared to Figure \ref{sketch}) of the tail, than shown in Figure \ref{sketch}, will result in a penumbral jet with right-handed untwisting.  

If the IRIS slit was placed along the jet then one could argue that the observed blueshift and redshift represent outflow from a reconnection site, situated in between, as proposed by \cite{kats07}. However, because we are looking at penumbral jets observed at a large angle to the IRIS slit, the observed blueshifts and redshifts are most probably representative of twisting motion of the magnetic field and plasma therein.   

Although we find clear signatures of twisting motion of the field in the spire of several large penumbral jets, not all penumbral jets display clear signature of twisting. Some of them have only redshift or only blueshift across them, thus, indicating some large penumbral jets might not twist. However, it is possible that we do not detect the other component, it being relatively very weak. \cite{depo14Science} have shown that swaying motions or field-aligned flows of modest size can make one side of the twist disappear \cite[see also,][]{depo12}.
 
The visibility of twisting in large penumbral jets also depends on which part (head, middle, tail) and phase (beginning, peak, decay) of the jet was covered by the IRIS slit. Evidently denser IRIS scans (e.g., in the second sunspot penumbra in our study, see Table \ref{t2}) increase the possibility of detecting twisting in large penumbral jets. Large penumbral jets seem to have a tendency of having stronger/dominant blueshift in the early phase and redshift in their last phase. This agrees with the idea presented in Figure \ref{sketch}. However some exceptions to this picture can be noticed in the two tables (see Tables \ref{t1} and \ref{t2}).

The visibility of only redshift or only blueshift across large penumbral jets might also depend on whether or not they are made by the eruption of a small-scale flux rope. Several recent studies on coronal jets in QRs, CHs and ARs have found that these have cool materials in them, in flux ropes, and are a result of minifilament eruptions \citep{ster15,pane16qr,pane17,ster16,ster17,pane18}. Due to much smaller sizes of penumbral jets, as compared to jets in QRs, CHs and ARs, whether there is an erupting ``microfilament" \citep[e.g.,][]{ster16moore} in these jets can not be seen. Moreover, although they display clear twisting in EUV images, smaller jets in CHs and QRs -- jetlets -- do not show any signatures of minifilament/cool materials inside them \citep{pane18a}. Whether large penumbral jets come from a microfilament eruption remains to be seen.

We are also not aware at this point if the visibility of blueshift/redshift in large penumbral jets varies from one sunspot penumbra to other sunspot penumbra, depending on their age, and activity level. The present work is the first report on the twisting motions of the magnetic field in large penumbral jets, and a more extensive work, both from observations and MHD simulations, is needed to supplement the findings in this paper. 

If penumbral jets form the way proposed by either \cite{kats07} or \cite{tiw16}, they may or may not twist, depending on the orientation and configuration of the magnetic field involved in the reconnection. If Alfv\'en waves are generated during magnetic reconnection driving a penumbral jet, one could expect to see twisting motions in the jet spire. Similarly, if large penumbral jets form the way jets in QRs, CHs or ARs do, by eruption of a minifilament flux rope, then we expect that they untwist/twist during the eruption. The fact that many of the large penumbral jets show twisting and many do not suggests that both types of mechanisms (convection-driven magnetic reconnection, and small-scale flux-rope eruption) could be at work in different penumbral jets. Further, we cannot rule out the possibility that the observed twisting in jet spires is caused by the propagation of Alfv\'en waves generated by convection-driven magnetic reconnection.  

\cite{este18} found weak line-of-sight (LOS) velocities ($<$5 \kms) inside penumbral microjets using \CaII\ data. Thus, penumbral jets might actually not be a site of strong field-aligned flows, but of modest flows. In that case the fast apparent motion of penumbral jets could be caused by a heating front (generated by magnetic reconnection at photospheric heights) moving much faster than the actual mass flow. This interpretation is similar to that of \cite{depo17} for network jets, which have much larger propagation speeds than Doppler shifts, and are proposed to be caused by heating fronts. Similar to large penumbral jets, network jets also show blue/red shifts, which are thought to be associated with twisting motions.

In the Dopplergram movies of both sunspot penumbrae there are some additional noticeable transient signals that somewhat resemble those that we have identified as large penumbral jets. For example in the movie Doppler\_Pen1.mp4 the raster starting at 15:02:03 UT displays a clear blueshift in two pixels (and a weak redshift in the next pixel down to the blueshifted pixels) along the first slit position, but no obvious jet-like activity in the corresponding slit-jaw (and the running difference) frames can be seen in the movie SJ\_Pen1.mp4. Because we do not find any jet-like activity in the slit-jaw images corresponding to some Doppler signals these Doppler signals could probably be related to a different type of activity, and not to large penumbral jets. We have ignored these Doppler signals in our present investigation.

Some locations of large penumbral jets remain bright for several minutes, displaying Ellerman bomb like signatures \citep[e.g.,][]{rutt13}. Whether large penumbral jets have any connection with Ellerman bombs remains to be seen. The most obvious similarities between Ellerman bombs and large penumbral jets are: (1) both show enhancements confined to the wings of the line profiles (c.f. \citealt{rear13}, and this study), (2) both show blue/red Doppler signal (often both) at their bases (c.f.  \citealt{wata11}, and this study), and (3) large penumbral jets and Ellerman bombs both form in the (higher) photosphere \citep{rutt13,tiw16}. The most obvious differences are: (1) Ellerman bombs are often found outside sunspots, whereas penumbral jets form inside the sunspot penumbra, (2) the average lifetime of 96 s of large penumbral jets, found in the present study, is much shorter than the lifetime of Ellerman bombs (average 566 s, c.f. \citealt{wata11}), (3) the extent/length of jets from Ellerman bomb locations are on average three times smaller than that of large penumbral jets, (4) the average extension and retraction speeds of Ellerman bombs (average 8 \kms, c.f. \citealt{wata11}) are much smaller than the average speed of large penumbral jets (36 \kms, this study), and (5) although large penumbral jets and Ellerman bombs both originate in the (higher) photosphere, Ellerman bombs do not have transition region signatures \citep{rutt13}, whereas large penumbral jets do show their direct response in the transition region \citep{viss15,tiw16}.

Likewise, whether moving bright penumbral dots \citep{tian14,alp16,deng16,sama17,youn18} have any connection with large penumbral jets needs further investigation.    

The thermal energy produced by each large penumbral jet ($3/2nk_BTV$) can be estimated to be $1.1\times10^{25}$ erg (for number density $n=10^{18} ~ m^{-3}$, Boltzmann constant $k_B=1.38\times10^{-23} ~ m^2 ~kg~ s^{-2} ~K^{-1}$, temperature $T=10^4~ K$, $V=3000 ~km\times (1350 ~km)^2$), which is about two orders of magnitude higher than that estimated for microjets \citep[\eg][]{kats07}. Therefore direct signatures of large penumbral jets in the transition region are not surprising \citep{tiw16}, and even a direct coronal response is plausible, and should be investigated. One must be careful with this interpretation however: if an energy of 10$^{25}$ erg is dumped very low in the atmosphere (e.g., in the upper photosphere), it is possible that much of the energy is lost to radiation and, thus, may not always lead to signatures in the transition region.
	
Because the changes in the energies, which are more meaningful, can not be calculated due to penumbral jets being transient, we make crude estimates of kinetic, potential (assuming solar gravitational force of 274 m s$^{-2}$) and total magnetic (with approximated magnetic field $B$ of 1 kG) energies of large penumbral jets, which come out to be of the order of 10$^{26}$, 10$^{25}$ and 10$^{29}$ erg, respectively, requiring only of order 0.1\%  of the total magnetic energy in the volume of a large penumbral jet be released to drive the jet.

This study of two sunspot penumbrae scanned by IRIS, one as a coarse 8-step raster, and the other as a medium dense 64-step raster, also highlights the importance of denser and faster rasters to address such science questions as those addressed in this research. To accurately measure different parameters, and estimate inaccuracies in our current measurements, e.g., relating to their lifetimes, sizes, twisting motions, blueshift/redshift at different parts and in different phases of jets, a much faster (higher-cadence) observation program is required that we plan to perform in future. 

\section{Conclusions}\label{conc}
From the IRIS spectra and slit-jaw observations of two sunspot penumbrae we have characterized large penumbral jets. We find evidence of twisting motion of magnetic field in the spire of several large penumbral jets. Hinode (SOT) observations show mixed-polarity magnetic field at the base of the most of large penumbral jets (with a crudely approximated flux cancellation rate of the order of 10$^{18}$ Mx $h^{-1}$). These two results, that large penumbral jets have mixed-polarity photospheric magnetic field and they twist, together may be compatible with a scenario in which large penumbral jets could be made the same way as EUV/X-ray coronal jets are made in CHs, QRs or surrounding ARs. To confirm this, higher resolution observations (e.g., from DKIST, and the future generation solar telescopes) and sophisticated MHD simulations are required to be performed in future. 

\acknowledgments
S.K.T. gratefully acknowledges support by NASA contracts NNG09FA40C (IRIS), and NNM07AA01C (Hinode). B.D.P. gratefully acknowledges support from NASA grants NNX16AG90G, and NNG09FA40C (IRIS). IRIS is a NASA small explorer mission developed and operated by LMSAL with mission operations executed at NASA Ames Research center and major contributions to downlink communications funded by ESA and the Norwegian  Space  Centre.  Hinode is a Japanese mission developed and launched by ISAS/JAXA, collaborating with NAOJ as a domestic partner, NASA and STFC (UK) as international partners. Scientific operation of the Hinode mission is conducted by the Hinode science team organized at ISAS/JAXA. This team mainly consists of scientists from institutes in the partner countries. Support for the post-launch operation is provided by JAXA and NAOJ (Japan), STFC (U.K.), NASA, ESA, and NSC (Norway). N.K.P’s research was supported by an appointment to NASA Postdoctoral Program at the  NASA MSFC, administered by Universities Space Research Association under contract with NASA. A.C.S and R.L.M acknowledge the support from the NASA HGI program, and by the Hinode Project. Some of the Hinode SOT/SP inversions conducted at NCAR under the framework of the Community Spectropolarimetric Analysis Center (CSAC; http://www.csac.hao.ucar.edu/) were used. This research has made use of NASA's Astrophysics Data System.




\end{document}